\documentstyle[12pt]{article}

\newcommand{\de}{\partial}
\newcommand{\be}{\begin{equation}}
\newcommand{\ee}{\end{equation}}
\newcommand{\bea}{\begin{eqnarray}}
\newcommand{\eea}{\end{eqnarray}}
\newcommand{\bd}{\begin{displaymath}}
\newcommand{\ed}{\end{displaymath}}

\def\fey{{\big / \kern-.80em D}}
\font\mybb=msbm10 at 12pt

\def\bb#1{\hbox{\mybb#1}}

\def\complex{{\bb{C}}}
\def\zet{{\bb{Z}}}
\def\real{{\bb{R}}}

\def\R4{\real^4}
\def\ie{{\it i.e.}}
\def\Tr{\mbox{Tr}}
\def\Ker{\mbox{Ker}}
\def\cf{{\cal F}}
\def\unita{{1 \kern-.30em 1}}
\def\cg{{\cal G}}
\begin{document}
\thispagestyle{empty}
\vskip 1.5cm
\begin{flushright}
ROM2F-96-32
\end{flushright}
\centerline{\large \bf Instanton Calculus and Nonperturbative } 
\centerline{\large \bf Relations in $N=2$ Supersymmetric 
Gauge Theories}
\vspace{2.2cm}
%%%%%%%%%%%%%%%%%%%%%%%%  AUTORI  %%%%%%%%%%%%%%%%%%%%%%%%%%%%%%
\centerline{\bf Francesco Fucito}
\vskip 0.1cm
\centerline{\sl Dipartimento di Fisica, 
Universit\`a di Roma II ``Tor Vergata"}  
\centerline{\sl I.N.F.N. \ -- \  Sezione di Roma II, }
\centerline{\sl Via della Ricerca Scientifica, \ \ 00133 \ Roma \ \ ITALY}
\vskip .3cm
\centerline{\bf Gabriele Travaglini}
\vskip 0.1cm
\centerline{\sl 
Dipartimento  di Fisica, Universit\`a di Roma ``La Sapienza"}
\centerline{\sl I.N.F.N. \ -- \  Sezione di Roma II, }
\centerline{\sl Via della Ricerca Scientifica, \ \ 00133 \ Roma \ \ ITALY}
\vskip 2.2cm
\centerline{\large \bf ABSTRACT}
{Using instanton calculus we check, in the weak coupling region,
the nonperturbative relation
$$ <\Tr\phi^2>=i\pi\left(\cf-{a\over 2}
{\partial\cf\over\partial a}\right)$$
obtained for a $N=2$ globally supersymmetric gauge theory.
Our computations are performed for instantons of winding
number $k$, up to $k=2$ and turn out to agree with previous 
nonperturbative results.}

\newpage
\setcounter{page}{1}
\setcounter{equation}{0}
\setcounter{section}{0}
\section{Introduction}

\indent 

In a recent work \cite{sw}, Seiberg and Witten have 
managed to compute
the quantum moduli space and the Wilsonian effective action
for the Yang--Mills theory with global $N=2$ supersymmetry
($N=2$ SYM from now on).
This achievement has been possible by using judiciously a 
certain number of educated guesses for the behavior of 
the moduli space of vacua of the theory and by exploiting
the unique properties of the $N=2$ SYM. In fact, the 
Wilsonian effective action of this theory, after having
used the Higgs mechanism, is completely determined once a
certain prepotential $\cf$ is known \cite{sei}. 
In turn, this prepotential
$\cf$ is determined if its global structure is known or 
postulated. This global structure is given by the monodromies
around the singular points of the prepotential $\cf$: the group
generated by these monodromies is a subgroup of $SL(2,\zet)$.

In the electric (or Higgs) phase of the theory the form of the
prepotential is known since, due to nonrenormalization 
theorems,
only the one loop term contributes at the perturbative level.
Moreover nonperturbative corrections due to instantons must be 
considered, leading to the final expression
\be
\cf (A)={1\over i \pi}\left(
{A^2 \over 2} \ln{ 2A^2\over\Lambda^2}-
\sum_{k=1}^\infty\cf_k{\Lambda^{4k}\over A^{4k-2}}\right) \ .
\label{f.1}
\ee
In (\ref{f.1}) $\Lambda$ is the renormalization group 
invariant scale and $A$ is a chiral superfield whose lowest
component squared is $a^2\equiv - 2u $,  \ie  \ 
it is the gauge invariant coordinate  
of the moduli space of vacua, when 
the gauge group is $SU(2)$ (which will be our choice from
now on), at least  for  large $u$ and $a$. 
The coefficients $\cf_k$  give the nonperturbative
contributions due to instantons. A formidable
check of the assumptions made in \cite{sw}, concerning
the symmetries of the moduli space, is thus given by matching
the coefficients $\cf_k$ against those obtained by instanton
calculus. This check must be performed in the weak 
coupling region in which instanton calculus can be reliably
performed. Some work has already been done along these 
lines. Nonperturbative
contributions induced by instantons can in fact be seen,
in the framework of  perturbation theory, as effective
four fermion vertices to be added to the tree level Lagrangian.
The computation of these 
effective vertices has already been performed
for the case of instantons of winding number one 
\cite{fp,is,yung} and two \cite{dkm}. 

The approach presented
in this paper  is somewhat different in that we will check 
the nonperturbative relation
\be
<\Tr\phi^2>=i\pi\left(\cf-{a\over 2}
{\partial\cf\over\partial a}\right)
\label{f.2}
\ee
found in \cite{mat}. 
Expanding the l.h.s. in (\ref{f.2}) as
\be \label{phisquare}
<\Tr\phi^2(a)>=- \frac{1}{2} a^2 - \sum_{k=1}^\infty \cg_k
{\Lambda^{4k}\over a^{4k-2}}
\label{f.3}
\ee
and substituting (\ref{f.1}) in (\ref{f.2}) we find
$\cg_k=2k\cf_k$ for a comparison with the results of 
\cite{klt}. The $\cg_k$'s can also be straightforwardly
checked against
the results of the recursion relation found in \cite{mat}.
%%%%%%%%%%%%%%%%%%%%%%%%%%%%%%%%%%%%%%%%%%%%%%%%%%%%%
\footnote{The reader should pay attention to different 
normalizations. The conventions of this letter for the $\cf_k$'s, 
which have opposite sign with respect to those of  
\cite{dkm},  are connected to \cite{mat,klt}
as: $\cf_k=-2^{6k-2}\cf_k^{KLT}=-i\pi 2^{2k}\cf_k^M$.}
%%%%%%%%%%%%%%%%%%%%%%%%%%%%%%%%%%%%%%%%%%%%%%%%%%%%% 

Supersymmetric instanton
calculus was developed  in two distinct ways
\cite{af,nsvz,akmrv} to study supersymmetry breaking. 
The main difference between these two 
approaches consists in giving or not 
an expectation value to the 
scalar (Higgs) field of the $N=2$ multiplet. Given the check
we want to perform the right choice is to follow \cite{af,nsvz,af2}
where such an expectation value for the scalar field is present.

This is the plan of the paper:
in section 2 we shall briefly discuss the 
basic ingredients of the Atiyah--Hitchin--Drinfeld--Manin (ADHM) 
construction of  instantons  which will be useful later on.
In section 3 we introduce the semiclassical expansion 
of Green functions in SUSY gauge theories, and check the relation
(\ref{f.2}) against a $k=1$ computation. In section 4 we  extend 
our considerations to a background of  Pontryagin  index 
$k=2$.

\section{A Brief Review of the ADHM Formalism}
\setcounter{equation}{0}
Before describing the actual computation we need to briefly discuss
the ADHM construction and collect some useful formulae.

As it is well--known, self--dual $SU(2)$ connections on $S^4$, can be 
put into one to one 
correspondence with holomorphic vector bundles of rank $2$ over  
$\complex P^3$
admitting a reduction of the structure group to its compact 
real form. The ADHM construction \cite{adhm,aty} gives all 
these holomorphic bundles and consequently all $SU(2)$ 
connections on $S^4$.
The construction is purely algebraic and we find it more 
convenient to use quaternionic notations. The points, 
$x$, of the one--dimensional 
quaternionic space $\bb{H}\equiv \complex^2\equiv\R4$ 
can be conveniently
represented in the form $x=x^\mu \sigma_\mu$, with 
$\sigma_\mu=(\unita, i\sigma_c), c=1,2,3.$ The $\sigma_c$'s 
are the usual Pauli matrices.
The conjugate of $x$ is $x^\dagger = 
x^\mu \sigma_\mu^\dagger$. 
A quaternion is said to be real if it is proportional to 
$\unita$ 
and imaginary if it has vanishing real part. 

The prescription to find an instanton of winding number $k$ is 
the following: introduce a $(k+1)\times k$ quaternionic matrix
linear in $x$
\be
\Delta=a+bx \  .
\label{f.4}
\ee
The (anti--hermitean) gauge connection is then 
written in the form 
\be
A^{cl}_\mu=U^\dagger\de_\mu U,
\label{trecinque}
\ee
where $U$ is a $(k+1)\times 1$ matrix of quaternions providing an
orthonormal frame of $\Ker \Delta^\dagger$. In formulae
\be
\Delta^\dagger U = 0,
\label{f.5}
\ee
\be
U^\dagger U =\unita_2,
\label{f.6}
\ee
where $\unita_2$ is the two--dimensional identity matrix. The 
constraint (\ref{f.6}) ensures that $A_\mu^{cl}$ 
is an element  of the Lie algebra of the $SU(2)$ gauge group.
The condition of self--duality on the field strength of 
(\ref{trecinque}) is imposed by restricting the matrix $\Delta$ 
to obey
\be
\Delta^\dagger\Delta=f^{-1}\otimes\unita_2,
\label{realita}
\ee 
with $f$ an invertible hermitean $k\times k$ matrix 
(of real numbers). 
In addition to the gauge freedom (right multiplication of $U$
by a unitary quaternion)  we have the freedom to perform 
the transformations
\be
\label{freedom}
\Delta\to Q\Delta R \ ,
\ee 
with $Q\in Sp(k+1), R\in GL(k,\real)$, which leave 
(\ref{trecinque}) invariant.

These symmetries  can be used to simplify the expressions
of $a$ and $b$. Exploiting this fact, in the
following we will 
choose the matrix $b$ to be
\be
b=-\pmatrix {0_{1\times k}\cr\unita_{k\times k}}.
\label{boh}
\ee

From (\ref{trecinque}), the field strength of the gauge field
can be computed and it is
\be
F_{\mu\nu}=2(U^\dagger bf\sigma_{\mu\nu}b^\dagger U),
\label{bohboh}
\ee
where $\sigma_{\mu\nu}=i\eta^a_{\mu\nu}\sigma^a$, with $\eta^a_{\mu\nu}$
the 't Hooft symbols.
Using (\ref{bohboh}) we also compute
\be
\Tr (F_{\mu\nu} F_{\mu\nu}) = 2 \Box
\Tr \left [ b^\dagger(1+P) b f \right],
\label{f.66}
\ee
where 
\be
P=UU^\dagger=1-\Delta f\Delta^\dagger
\label{f.666}
\ee
is the projector on the kernel of 
$\Delta^\dagger$.
(\ref{f.66}) will turn out to be important in the following.

The bosonic zero--modes, $Z_{\mu}$, of the gauge--fixed
second order differential  operator
\be
M_{\mu \nu} =- D^{2} (A^{cl})\delta_{\mu\nu}
-2  F^{cl}_{\mu \nu} ,
\label{f.7}
\ee
which describes the quantum fluctuations of the gauge fields,
can be found by noting that the transverse 
fluctuations
of a self--dual configuration must   satisfy the relations 
\be
{}^{\ast}( D_{[\mu} Z_{\nu ]} ) = D_{[\mu} Z_{\nu ]},\quad 
 D_{\mu} Z_{\mu}=0 . 
\label{f.8}
\ee
This allows to write $Z_{\mu}$ in terms of the quantities
appearing in  (\ref{f.4}), as \cite{osb}
\be
Z_\mu=U^\dagger C\bar\sigma_\mu f b^\dagger U-
U^\dagger bf\sigma_\mu C^\dagger U.
\label{f.9}
\ee
For (\ref{f.9}) to satisfy (\ref{f.8}), the $(k+1)\times k$ 
matrix $C$ must obey
\be
\Delta^\dagger C=(\Delta^\dagger C)^T,
\label{f.10}
\ee
where the superscript $T$ stands for transposition of the
quaternionic
elements of the matrix (without transposing the quaternions 
themselves).
In our case, the number of independent $Z_{\mu}$ is $8k$ (the
dimension of the moduli space of the instanton). $C$ has 
$k(k+1)$ quaternionic elements, which are subject to the
$4k(k-1)$ constraints (\ref{f.10}). The number of $C$'s 
satisfying (\ref{f.10}) is thus $8k$ as desired. This is the
reason why in the following we will sometimes attach a subscript
$r=1,\ldots,8k$ to the zero--modes. 

Fermionic zero--modes
are easily deduced from (\ref{f.9}) by remarking that, 
due to $N=2$ SUSY,
\be 
\lambda_{\beta \dot{A}}^{(r)} = 
\sigma^{\mu}_{\beta \dot{A}} Z_{\mu}^{(r)} ,
\label{dual}
\ee
where  $\dot{A}=1,2$ labels the two SUSY charges.
Furthermore, the superposition 
 of the $Z_{\mu}^{(r)}$ was computed in 
\cite{osb} to be
\be
(Z^{(r)},Z^{(s)}) = {4\pi^2  \over g^2}
\Tr\left[C^\dagger_r (1+P_\infty)
C_s\right] \equiv{4\pi^2  \over g^2}
(C_r , C_s ) \ ,
\label{f.11}
\ee
where $P_\infty = 1- b b^{\dagger}$ 
is the projector $P$ evaluated in the limit
$|x|\to\infty$. The superpositions  of the bosonic zero--modes 
are then tied to
the Jacobian which yields the integration measure for the
bosonic collective coordinates. This integration measure
is easily written once the variations of the bosonic fields
with respect to the instanton moduli are known. But the 
zero--modes (\ref{f.9}) are also transverse 
to allow the factorization of the infinite volume of 
the gauge group via the introduction
of the Faddeev--Popov determinant \cite{yar}. It is thus useful
to separate, in the expression of the bosonic zero--modes, 
the variation with respect to the instanton moduli from
the gauge part needed to make it transverse \cite{th,ber}.
In the ADHM formalism this yields to the formula for the 
variation of the gauge connection \cite{cgot}
\be
\delta_rA_\mu=U^\dagger (\delta_r\Delta )\bar\sigma_\mu 
f b^\dagger U-
U^\dagger bf\sigma_\mu (\delta_r\Delta)^\dagger U+
[D_\mu,\delta g]
\label{f.12}
\ee
where $\delta_r$, $r=1,\ldots,8k$ stand for the variations with
respect to the instanton moduli, and $\delta g$, satisfying 
$\delta g+\delta g^\dagger=0$, is an arbitrary infinitesimal
gauge transformation. The unconstrained
variations $\delta_r\Delta$, which give the integration
over the collective coordinates, cannot be easily traded
with (\ref{f.11}) since the $C$'s appearing in that formula
are constrained by (\ref{f.10}). The complete relation between 
the $\delta_r\Delta$ and $C$'s is given by 
\be
(C_r,C_s)=(\delta_r\Delta,\delta_s\Delta)
+2K_{rji}M_{ij,lm}K_{slm}.
\label{f.13}
\ee
The explicit expression of the matrices $K, M$, 
which parametrize the freedom to transform the ADHM data 
as in (\ref{freedom}),  can be found in \cite{osb}.

\section{The k=1 Semiclassical Computation}
\setcounter{equation}{0}
We  now briefly review the strategy 
to perform 
semiclassical computations in supersymmetric gauge theories,
in the context of the constrained instanton method.

The classical potential for the complex scalar field 
$\phi$ of an $N=2 $ SYM gauge theory
\be
V_D={1\over 2}( \epsilon^{abc} \phi^{b}
\phi^{\dagger c})^{2}
\label{pot}
\ee
has flat directions  when $\phi $  is a $SU(2)$  gauge 
transform of $\phi = a^{c} \sigma_{c} / 2i $, where 
$a^{c} = a \delta^{c3}$ and $a$ is  a  complex number.  
Following 't Hooft \cite{th}, 
we shall then expand the action functional 
around  a properly chosen field configuration, which is the 
solution of the  equations 
\be
D_{\mu} (A) F_{\mu \nu} = 0 \ \ ,
\ee
\be\label{classphi}
D^{2} (A) \phi_{\rm cl} = 0 \ \ , \ \ \lim_{|x| 
\rightarrow \infty}
\phi_{\rm cl} \equiv \phi_\infty = a \frac{\sigma_{3}}{2i}. 
\label{finfty}
\ee
The first equation admits instantonic solutions.
When $a=0$, \mbox{(\ref{classphi})} admits the 
trivial solution 
$\phi =0$ only.  When $a \neq 0$, we shall decompose
the fields $\phi$, $\phi^{\dagger}$ as
\be
\begin{array}{c c}
\phi= \phi_{\rm cl} + \phi_{Q} \ \ , & 
\ \ \phi^{\dagger}= (\phi_{\rm cl})^{\dagger}
+ \phi_{Q}^{\dagger} \ \ ,
\end{array}
\ee
and integrate over the quantum fluctuations $\phi_{Q}$, 
$\phi_{Q}^{\dagger}$.

The integration over bosonic zero--modes can be traded with 
an integration over collective coordinates, at the cost of 
introducing  the corresponding jacobian.
The existence of fermion zero--modes  is the 
way by which Ward identities  related with the 
group of   chiral symmetries of the theory,
come into play.  
When $a=0$,  the  anomalous  $U_{R}(1)$ symmetry
\be
\begin{array}{c c}
\lambda \longrightarrow e^{i \alpha} \lambda \ \ ,
& \ \ 
\phi \longrightarrow e^{2 i \alpha}  \phi
\end{array}
\ee
and gauge invariance  allow 
a nonzero result    for the 
Green functions   with $n$ insertions of
the gauge invariant quantity  $(\phi^{a } \phi^{a })(x)$ only 
when $n = 2 k $.
These correlators possess the right operator insertions 
needed   to saturate 
the integration over the  Grassmann parameters, and 
due to  supersymmetry,  they  are also 
position independent.  
On the other hand when  $a\neq 0$  
the correlator $\left\langle \phi^{a} \phi^{a} \right\rangle$
has a complete expansion in terms of instanton contributions,
as  in   \mbox{(\ref{phisquare})}.  
The $k=1$ term has been computed in  \cite{FS}
in the framework of the constrained superinstanton  formalism 
of \cite{nsvz}.
In the following we will use the component formalism,
which we find simpler, to compute 
the $k=1, 2$  coefficients  of the instanton expansion. 
%%%%%%%%%%%%%%%%%%%%%%%%%%%%%%%%%%%%%%%%
%%%%%%%%%%%%%%%%%%%%%%%%%%%%%%%%%%%%%%%%
%%%%%%%%%    k=1   %%%%%%%%%%%%%%%%%%%%% 

Fermion zero--modes are found by solving the equation 
\be 
D_{\mu} \bar{\sigma}^{\dot{\alpha}\beta}_{\mu} 
\lambda_{\beta \dot{A}} = 0 \ \ .
\label{g.4}
\ee
For instantons of winding number $k=1$
 the whole set of solution  
of this equation is obtained via SUSY and superconformal 
transformation, which yield 
\be
\lambda_{\alpha \dot{A}}^{a} = 
\frac{1}{2}F_{\mu \nu}^{a}
(\sigma_{\mu \nu})_{\alpha}^{\ \beta}
 \zeta_{\beta \dot{A}} \ \ ,
\label{g.5}
\ee
where 
$\zeta = \xi + (x-x_{0})_{\mu} \sigma_{\mu} 
\bar{\eta}/\sqrt{2} \rho $, 
$\xi$, $\bar{\eta}$ being two arbitrary quaternions
of Grassmann numbers.  
It is instructive, and useful for the computations to be performed in
the next chapter, to deduce this result using the ADHM expressions 
(\ref{f.9}), (\ref{f.10}), (\ref{dual}). 
For $k=1$ the constraint (\ref{f.10}) is always satisfied since 
$\Delta^\dagger C$ is a single quaternion. 
Given a matrix 
\be
\Delta=\pmatrix{v\cr x_0-x},
\label{g.50}
\ee
and choosing $C$ to be 
\be
C=\pmatrix{\sqrt{2} C_0 \cr 2C_1},
\label{g.51}
\ee
with $C_0, C_1$ two arbitrary quaternions,
we can substitute (\ref{g.51}) in (\ref{f.9}) 
to find, using (\ref{bohboh})
\be 
Z_\mu= 2 F_{\mu\nu}B_\nu,
\label{g.52}
\ee
where
\be
B=B_\nu\sigma^\nu=C_1 + (x-x_{0}) {\bar{v} \over v^2}  
{C_0\over\sqrt{2}}.
\label{g.53}
\ee
Interpreting $C_0, C_1$ as Grassmann variables
and using (\ref{dual}) we obtain (\ref{g.5}) with $B$ replacing 
$\zeta$.  

The correct fermionic integration measure is given 
by the inverse of the determinant of the matrix whose 
entries are the scalar products of the fermionic zero--mode
eigenfunctions. This scalar product is induced by the 
kinetic terms in the action and for arbitrary $SU(2)$
valued spinors $f,g$ is
\be
(f,g)\equiv \sum_{a=1}^3\int d^4x \ (f_\alpha^a)^\ast
\bar\sigma_0^{\dot\alpha\beta}g^a_\beta.
\label{norm}
\ee
The fermionic measure now reads
\be
d^{4} \xi d^{4} \bar{\eta} \left( \frac{g^2}{32 \pi^2} \right)^{4}
 \ \ ,
\ee
where $d^{4} \xi d^{4} \bar{\eta} \equiv
d^{2} \xi_{\dot{1}} d^{2} \xi_{\dot{2}}
d^{2} \bar{\eta}_{\dot{1}}  d^{2} \bar{\eta}_{\dot{2}}  $.
Once auxiliary fields are eliminated, the action is
\be
S =S_{\rm G} +  S_{\rm H} +  S_{\rm F} + S_{\rm Y}+ S_{\rm D}. 
\label{azione}
\ee 
$S_{\rm G}$ is the usual gauge field action, 
$S_{\rm F}[\lambda,\bar\lambda,A]= \int d^{4}x\ 
\bar{\lambda}^{\dot{A} a } \Bigl[ \ 
\fey(A)\lambda_{\dot{A}} \Bigr]^{a}$ 
and
$S_{\rm H}[\phi,\phi^\dagger,A]=\int d^{4}x\ (D \phi)^{\dagger a}
(D \phi)^{a}$   
are the kinetic terms for the Fermi and Bose fields  
minimally coupled to the gauge field $A_{\mu}$.  
The Yukawa interactions are given by
\be
S_{\rm Y} [ \phi , \phi^{\dagger} , \lambda , \bar{\lambda}] = 
\sqrt{2} g  \epsilon^{abc} 
\int d^4x \ \phi^{a \dagger} (\lambda_{\dot{1}}^{b} 
\lambda_{\dot{2}}^{c}) \ + \ {\rm h.c.}
\ee
and finally
$ S_{\rm D} =  \int d^{4} x\  V_D$
comes from the potential term  (\ref{pot})  for the complex 
scalar field $\phi$.

The evaluation of the correlator
$<\phi^{a} \phi^{a} >$ in the semiclassical approximation 
around an instantonic background of winding number $k=1$ 
yields 
\bea \label{k=1}
&&<\phi^{a} \phi^{a} > = 
\nonumber \\
&&\int d^{4}   x_{0} d \rho \ 
\left( \mu^8 \frac{2^{10} \pi^{6} \rho^{3}}{g^{8}} \right)
e^{-\frac{8 \pi^2}{g^2} - 4 \pi^{2} |a|^{2} \rho^{2}}    
\int [ \delta Q  \delta \lambda ] \delta \bar{\lambda}  
\delta \phi^{\dagger}_{Q} \delta \phi_{Q}
\delta \bar c  \delta c  
\nonumber \\
&&
\exp \biggl[ - S_H[\phi_Q,\phi_Q^\dagger,A^{cl}]-S_F[\lambda,
\bar\lambda,A^{cl}]+
\nonumber \\
&& - \frac{1}{2} \int d^4x \ Q_{\mu} M_{\mu \nu} Q_{\nu}
- \int d^4x \ \bar c D^2 (A^{cl}) c \biggr]
\nonumber \\
&&\mu^{-\frac{1}{2} (4+4) }\int
d^{4} \xi d^{4} \bar{\eta} 
\left( \frac{g^2}{32 \pi^2} \right)^{4}
\
\exp \left[ -S_{\rm Y} [ \phi_{\rm cl} + \phi_{Q}, 
 (\phi_{\rm cl})^{\dagger} + \phi_{Q}^{\dagger}  , 
\lambda^{(0)},
\bar\lambda=0 ] 
\right] 
\nonumber \\ 
&&(\phi_{\rm cl} + \phi_{Q})^{a} ( \phi_{\rm cl} + 
\phi_{Q})^{a}(x) \ \ \ .
\eea
Let us now explain where 
the different terms in \mbox{(\ref{k=1})}
come from: 
\begin{itemize}
\item[{\bf 1.}]
$ d^{4}   x_{0} d \rho \ 
\left( \mu^8 \frac{2^{10} \pi^{6} \rho^{3}}{g^{8}} \right)$
is the bosonic measure \cite{th,ber} after the integration over
$SU(2) / \zet_{2}$ global rotations in color space has been
performed.  $x_{0}$ and 
$\rho$ are the center and the size of the instanton
(see  (\ref{g.50}),  with  $\rho \equiv |v|$).
\item[{\bf 2.}]
$S_H[\phi_{cl},(\phi_{cl})^\dagger,A^{cl}]=4 \pi^{2} |a|^{2} 
\rho^{2}$,
is the contribution of the classical Higgs action,
and has been   computed by 't Hooft \cite{th}.
\item[{\bf 3.}]
The second line  include the quadratic
approximation of the different 
kinetic operators for the quantum fluctuation 
of the fields and the 
symbol $[\delta \lambda \delta Q]$ denotes integration 
over nonzero--modes.  $\bar c$ and $c$ are the usual ghost fields, 
$\int d^4x \ \bar c D^2 (A^{cl}) c$ 
being the corresponding term in the action.

\item[{\bf 4.}]
$S_{\rm Y} \left[ \phi ,  \phi^{\dagger}  , \lambda^{(0)},
\bar\lambda=0 \right] $
is the Yukawa action calculated with the complete expansion 
of the fermionic  fields replaced by their   
projection over the zero--mode 
subspace. According to the
index theorem for the Dirac operator in the background of a 
self--dual gauge field configuration, 
we have only zero--modes of one
chirality, so that this term reduces to 
$\sqrt{2} g  \epsilon^{abc} 
\int \phi^{a \dagger} (\lambda_{\dot{1}}^{(0) b} 
\lambda_{\dot{2}}^{(0) c})$.
\item[{\bf  5.}]
$\mu^{8-\frac{1}{2}(4+4)} e^{- \frac{8 \pi^2}{g^2}}
= \Lambda^{4}$, where $\Lambda$  is
the (one loop) $N=2$  SYM renormalization group invariant scale
with gauge group $SU(2)$. $\mu$ comes from the Pauli--Villars
regularization of the determinants and the exponent is
$b_1k=(n_B-n_F/2)$ where $n_B,n_F,b_1$ are the number of bosonic,
fermionic zero--modes and the first coefficient of the 
$\beta$--function of the theory. 
\end{itemize}
After the  integration over $\phi$, $\phi^{\dagger}$ and
 the nonzero--modes,  the $\phi_{Q}$   insertions
get replaced by $\phi_{\rm inh}$, where
\be 
\phi_{\rm inh}^{a}= \sqrt{2} g  \epsilon^{bdc} 
[( D^{2})^{-1}]^{ab} (\lambda_{\dot{1}}^{(0) d}  
\lambda_{\dot{2}}^{(0) c}) \ \ ,
\label{fiino}
\ee
and the determinants of the various kinetic operators cancel
against each  other \cite{divecchia}.
The r.h.s. of \mbox{(\ref{k=1})} now reads 
\bea
&&\Lambda^{4}\int d^{4}   x_{0} d \rho \ 
\left(\frac{2^{10} \pi^{6} \rho^{3}}{g^{8}} \right)
e^{- 4 \pi^{2} |a|^{2} \rho^{2}}    
\nonumber \\
&&\int d^{4} \xi d^{4} \bar{\eta}
 \left( \frac{g^2}{32 \pi^2} \right)^{4}
 \
\exp \left[ - \sqrt{2} g  \epsilon^{abc} 
\int \phi_{cl}^{\dagger} (\lambda_{\dot{1}}^{(0) b} 
\lambda_{\dot{2}}^{(0) c}) \right] 
\nonumber \\ 
&&(\phi_{\rm cl} + \phi_{\rm inh })^{a} 
 ( \phi_{\rm cl} + \phi_{\rm inh})^{a}(x)
 \ \ \ .
\label{correlatore}
\eea
A straightforward calculation shows that 
\cite{nsvz2}
 \be
S_{\rm Y} \left[ \phi_{cl} ,  \phi_{cl}^{\dagger}  , 
\lambda^{(0)},
\bar\lambda=0  \right] = 
\frac{(a^c)^{\ast} g}{\sqrt{2}}
 \left( \bar{\eta}_{\dot{1}} \sigma^{c} 
\bar{\eta}_{\dot{2}} \right)
 \left( \frac{g^2}{32 \pi^2} \right)^{-1} \ \ .
\ee
Moreover it is also  easy to convince  oneself that 
(\ref{fiino}) is solved by
\be
\phi_{\rm inh }^{a} = \sqrt{2} ( \zeta_{\dot{1}} 
\lambda_{\dot{2}}^{a} ) \ \ ,
\ee
as it can be checked by substituting in
\be
[ D^{2}]^{ab} \phi_{\rm inh}^{b}= \sqrt{2} g  \epsilon^{abc} 
(\lambda_{\dot{1}}^{(0) b }  
\lambda_{\dot{2}}^{(0) c}) \ \ .
\label{fiino1}
\ee

The Yukawa action does {\em not} contain  
the Grassmann parameters of the zero--modes coming
from   SUSY transformations. 
As a consequence 
the only 
nonzero contributions are  obtained by  picking out
the terms in the $\phi_{\rm inh}$ insertions 
which contain the SUSY solutions
of the Dirac equation. We thus completely
disregard the $\phi_{\rm cl}$ pieces. 
Since  
\be
\phi_{\rm inh }^{a} \phi_{\rm inh }^{a} =
 - \zeta_{\dot{1}}^{2}
 (\lambda_{\dot{2}}^{a}\lambda_{\dot{2}}^{a})
= - \zeta_{\dot{1}}^{2} \zeta_{\dot{2}}^{2} 
(F_{\mu \nu}^{a} F_{\mu \nu}^{a}) \ \ ,
\ee
this amounts to say 
\be
(\phi_{\rm cl} + \phi_{\rm inh })^{a}  
( \phi_{\rm cl} + \phi_{\rm inh})^{a}
\longrightarrow
 - \xi_{\dot{1}}^{2} \xi_{\dot{2}}^{2} 
(F_{\mu \nu}^{a} F_{\mu \nu}^{a})  .
\label{fireplace}
\ee
The integration over non SUSY zero--modes is then 
dealt with by performing the integral
\be
\int d^{4} \bar{\eta} \left( \frac{g^2}{32 \pi^2} \right)^{2} \ 
\exp \left[ 
\frac{(a^c)^{\ast} g}{\sqrt{2}} \left( \bar{\eta}_{\dot{1}} \sigma^{c} 
\bar{\eta}_{\dot{2}} \right) \left( \frac{g^2}{32 \pi^2} \right)^{-1} 
\right]=\frac{g^2}{2} (a^{\ast})^{2}.
\label{intsusy}
\ee
(\ref{k=1}) now becomes
\bea
<\phi^{a} \phi^{a} > \ \ = 
&& \Lambda^{4}\int d^{4}   x_{0} d \rho \ 
\left(\frac{2^{10} \pi^{6} \rho^{3}}{g^{8}} \right)
e^{- 4 \pi^{2} |a|^{2} \rho^{2}}    
 (F_{\mu \nu}^{a} F_{\mu \nu}^{a})  
\nonumber \\
&&\frac{g^2}{2} (a^{\ast})^{2}
\int d^{4} \xi   \left( \frac{g^2}{32 \pi^2} \right)^{2} \ 
\xi_{\dot{1}}^{2} \xi_{\dot{2}}^{2} 
 \ \ \ .
\label{g1}
\eea
A simple computation using the explicit form of the $k=1$
gauge connection, shows that $ F_{\mu \nu}^{a} F_{\mu \nu}^{a}$
is 
a function of the difference
$x-x_{0}$.
We can then immediately integrate over
$x_{0}$ remembering that 
$\int d^{4} x \  F_{\mu \nu}^{a} F_{\mu \nu}^{a} = 
32 \pi^2 / g^2$. 
The remaining integrations over $\xi$ and
$\rho$ in (\ref{g1}) are trivial and yield
\cite{fp}
\be
<\phi^{a} \phi^{a} > = \frac{2}{g^4}\frac{\Lambda^{4} }{a^2} \ \ .
\ee
This result agrees with the coefficient $\cg_1$ found in \cite{mat,klt}.
%%%%%%%%%%%%%%%%%%%  k=2  %%%%%%%%%%%%%%%%%%%%%%%
\section{The k=2 Computation}
\setcounter{equation}{0}
We now come to the $k=2$ computation.
There are several modifications to take into account, 
with respect to the $k=1$ case
but the general strategy is unchanged.
We start by giving the form of the matrix $\Delta$
of (\ref{f.4}) 
\be
\Delta=\pmatrix{v_1 & v_2\cr x_1 - x & d\cr d & x_2 - x}
=\pmatrix{v_1 & v_2\cr e & d\cr d & -e}+b(x-x_0) \ .
\label{f.14}
\ee
The constraint (\ref{realita}) is obeyed if 
\be
d = {1 \over 2} {z \over z^2}  ( \bar{v}_{2} v_1 - 
\bar{v}_{1} v_2 ),
\label{f.15}
\ee 
with $ z =x_1-x_2$ \cite{csw}. We find it more convenient
to expose the role of the center of the instanton, the part
proportional to the matrix $b$ of (\ref{boh}), because this
will be central in the integration which we will perform later.
This is achieved with the substitutions $x_0=(x_1+x_2)/2, e=
(x_1-x_2)/2$ which gives the other form of the matrix 
$\Delta$ in (\ref{f.14}).

We also need the form of the matrix $C$ appearing in (\ref{dual})
%%%%%%%%%%%%%%%%%%%%%%%%%%%
\footnote{The elements of this matrix must be 
interpreted as Grassmann numbers from the point of view of
the functional integration.}
%%%%%%%%%%%%%%%%%%%%%%%%%%%%%%%%%
which is constrained by (\ref{f.10}). Since this constraint 
is very similar to (\ref{realita}) (to get convinced of 
this fact
just think that two solutions of (\ref{f.10}) are given 
by $C=a,b$)
it is convenient to choose a form of $C$ which parallels 
(\ref{f.14})
\be
C=\pmatrix{\nu_1 & \nu_2\cr \xi_1 & \delta\cr \delta & \xi_2}
=\pmatrix{\nu_1 & \nu_2\cr \eta & \delta\cr 
\delta & -\eta}- b\xi_0.
\label{f.155}
\ee
The constraint (\ref{f.10}) is satisfied imposing
\be
\delta={z\over z^2}[2\bar d\eta+\bar v_2\nu_1-\bar v_1\nu_2].
\label{f.16}
\ee

In analogy with the points 2, 4 of the previous calculation 
for the $k=1$ case, we have to compute\footnote{The 
following expression contains the prescription for
computing $S_H + S_Y$ on the saddle point for which 
the only zero--modes are the left--handed ones.}
\be
S_H[\phi_{cl}+\phi_{inh},(\phi_{cl})^\dagger,A^{cl}]
+S_Y[\phi=0,(\phi_{cl})^\dagger,\lambda^{(0)},\bar\lambda=0].
\label{f.166}
\ee
This computation involves only
the contributions of the $\phi_{cl}$
function and  of the $\phi_{inh}$ field 
at the boundary of the physical space. It has 
been performed in \cite{dkm} and it yields
\bea
&&S_H+S_Y=4\pi^2|a|^2(|v_1|^2+|v_2|^2)-4\pi^2 {
[\Tr(v_1\bar v_2-v_2 \bar v_1)\phi_\infty]^2\over 
|v_1|^2+|v_2|^2+4(|d|^2+|c|^2)}
\nonumber\\
&&+2\sqrt{2}\pi^2\epsilon^{\dot A\dot B}
\epsilon^{\alpha\gamma}\biggl[(\nu_i)_{\gamma\dot A}
(\phi_\infty)_\alpha{}^\beta(\nu_i)_{\beta\dot B}+
\biggl({\Tr
(v_1\bar v_2-v_2 \bar v_1)\phi_\infty\over 
|v_1|^2+|v_2|^2+4(|d|^2+|c|^2)}\biggr)\nonumber\\
&&((\nu_1)_{\gamma\dot A}
(\nu_2)_{\alpha\dot B}+2\eta_{\gamma\dot A}
\delta_{\alpha\dot B})\biggr] \ \ ,
\label{f.17}
\eea
where $\phi_\infty$ was defined in (\ref{finfty}). 

Let us comment on (\ref{f.17}):
in the Yukawa action the variable $\xi_0$ is missing.
In fact, the expectation value of the scalar field has broken
the conformal but not the translational invariance of  the  action.
The effect is the appearance, in the action, of the 
collective coordinates related to these symmetries.
As SUSY is still a symmetry, the collective coordinates
associated to it must be missing in (\ref{f.17}), 
which is what we observe.
In complete analogy with the $k=1$ case, 
the Grassmann parameters can now be divided into two sets:
those which do not appear in the action
(connected to SUSY) must be isolated in the $\phi_{inh}$
piece to cancel against the measure.
Those which appear in the action, will not appear in the
insertion of $\phi^a\phi^a$: all fermionic zero--modes 
are lifted but the eight SUSY ones. There is another 
consequence of this observation. On the r.h.s. of (\ref{fiino})
there are fermionic fields expanded in the basis of the 
zero--modes given by (\ref{dual}). See also the $k=1$ case 
(\ref{g.52}), (\ref{g.53}) for a comparison. 
The previous observation thus 
suggests us to solve (\ref{fiino1}) only for those fermionic fields
containing the Grassmann parameters related to 
SUSY transformations. As a consequence (\ref{fireplace}) 
still holds. The term $F_{\mu \nu}^{a} F_{\mu \nu}^{a}$
is given by (\ref{f.66}), and in (\ref{f.14}) $\Delta$ was 
parametrized in order to be a function of $x-x_0$.
Since (\ref{f.66}) depends only on $\Delta$,
$F_{\mu \nu}^{a} F_{\mu \nu}^{a}$ is a function of $x-x_0$ too.
 
Given all these  observations we can write the correlator
for the case $k=2$ as 
\bea
<\phi^{a} \phi^{a} > &=& {\Lambda^8\over {\cal S}}\int d^4e 
d^4v_1d^4v_2 d^4\eta
d^4\nu_1d^4\nu_2\left({J_{Bose}\over 
J_{Fermi}}\right)^{1\over 2}e^{-S_H-S_Y}\nonumber\\
&& \int d^4\xi_0
(\xi_0)_{\dot 1}^2(\xi_0)_{\dot 2}^2
\int d^4x_0 \ F_{\mu \nu}^{a} F_{\mu \nu}^{a},
\label{f.18}
\eea
where ${\cal S}=16$ is a statistical weight computed in \cite{osb}.
The integrations in the last line of (\ref{f.18}) can be performed
immediately after trading the $x_0$ with the $x$ integration by 
shifting variable, and give 
\be
\int d^4\xi_0
[(\xi_0)_{\dot 1}]^2[(\xi_0)_{\dot 2}]^2
\int d^4x_0 \ F_{\mu \nu}^{a} F_{\mu \nu}^{a}=4 \cdot 
{64\pi^2\over g^2}.
\label{inttrans}
\ee

The fermionic Jacobian, $J_{Fermi}$, is obtained from (\ref{dual}),
(\ref{f.11}) while $J_{Bose}$ was computed in \cite{osb, dkm}.
Putting these results together we get
\be
\left({J_{Bose}\over J_{Fermi}}\right)^{1\over 2}={2^{10}\over \pi^8}
{||e|^2-|d|^2|\over |v_1|^2+|v_2|^2+4(|d|^2+|c|^2)} .
\label{jac}
\ee

After substituting (\ref{jac}) in (\ref{f.18}), the remaining 
integrations can be performed and 
give $5/(32\pi^2 a^6 g^6) \cite{dkm}$.

Collecting all these results we finally find
\be
<\Tr\phi^2>=-{5\over 4}{\Lambda^8\over g^8 a^6},
\ee
which is in agreement with the results of \cite{mat,klt}

\vskip 1.5cm

\leftline{\bf\large Acknowledgments}
\vskip .5cm
The authors would like to thank
Massimo Bianchi, Marco Matone and Gian Carlo Rossi for
interesting conversations.


\newpage
\begin{thebibliography}{99}

\bibitem{sw}
{N.~Seiberg and E.~Witten, Nucl. Phys. {\bf B426} (1994) 19;
{\it ibid.} {\bf B431} (1994) 484.}

\bibitem{sei}
{N.~Seiberg, Phys. Lett. {\bf 206B} (1988) 75.}


\bibitem{fp}
{D.~Finnell and P.~Pouliot, Nucl. Phys. {\bf B453} (1995) 225.}

\bibitem{is}
{K.~Ito and N.~Sasakura, {\it One-Instanton Calculations
in $N=2$ Supersymmetric $SU(N_c)$ Yang--Mills Theory}, preprint
KEK-TH-470, {\tt hep-th/9602073}.}

\bibitem{yung}
{A.~Yung, {\it Instantons--induced Effective Lagrangian in the 
Seiberg--Witten Model}, preprint SWAT/96/111, 
{\tt hep-th/9605096}.}

\bibitem{dkm}
{N.~Dorey, V.~Khoze and M.~Mattis, {\it Multi--Instanton Calculus
in $N=2$ Supersymmetric Gauge Theories}, 
{\tt hep-th/9603136}.}

\bibitem{mat}
{M.~Matone, Phys. Lett. {\bf 357B} (1995) 342.}

\bibitem{klt}
{A.~Klemm, W.~Lerche and S.~Theisen, {\it Nonperturbative 
Effective
Actions of $N=2$ Supersymmetric Gauge Theories}, preprint 
CERN-TH/95-104, LMU-TPW 95-7 and 
{\tt hep-th/9505150}.}

\bibitem{af}
{I.~Affleck, M.~Dine and N.~Seiberg, Phys. Rev. Lett. 
{\bf 51} (1983) 1026;  Nucl. Phys. 
{\bf B241} (1984) 493.}

\bibitem{nsvz}
{V.~Novikov, M.~Shifman, A.~Vainshtein and 
V.~Zakharov, Nucl. Phys. {\bf B260} (1985) 157.}

\bibitem{akmrv}
{G.C.~Rossi and G.~Veneziano, Phys. Lett. {\bf B138} (1984) 195;
D.~Amati, K.~Konishi, Y.~Meurice, G.C.~Rossi and 
G.~Veneziano, Phys. Rep. {\bf 162C} (1988) 169.} 

\bibitem{af2}
{I.~Affleck, Nucl. Phys. {\bf B191} (1981) 429.}

\bibitem{adhm}{M.~Atiyah, V.~Drinfeld, N.~Hitchin and Yu.~Manin, 
Phys. Lett. 
{\bf 65A} (1978) 185.}

\bibitem{aty}{M.~Atiyah, {\it Geometry of Yang--Mills 
fields}, Lezioni 
Fermiane, Accademia Nazionale dei Lincei e Scuola Normale 
Superiore, Pisa 
(1979).}

\bibitem{osb}
{H.~Osborn, Ann. Phys. {\bf 135} (1981) 373.}

\bibitem{yar}
{D.~Amati and A.~Rouet, Nuovo Cimento {\bf 50A} (1979) 265;
L.~Yaffe, Nucl. Phys. {\bf B151} (1979) 247.}

\bibitem{th}
{G.~'t~Hooft, Phys. Rev. {\bf D14} (1976) 3432.}


\bibitem{ber}
{C.~Bernard, Phys. Rev. {\bf D19} (1979) 3013.}


\bibitem{cgot}
{E.~Corrigan, P.~Goddard, H.~Osborn and S.~Templeton,
Nucl. Phys. {\bf B159} (1979) 469.}


\bibitem{FS}
{J.~Fuchs and M.~Schmidt,
Z. Phys. {\bf C30} (1986)  161.}

\bibitem{divecchia}
{A.~D'Adda  and P.~Di Vecchia, Phys. Lett. 
{\bf 73B} (1978) 162.}

\bibitem{nsvz2}
{V.~Novikov, M.~Shifman, A.~Vainshtein and 
V.~Zakharov, Nucl. Phys. {\bf B223} (1983) 445 .}

\bibitem{csw}{N.~Christ, N.~Stanton and  E.~Weinberg, 
 Phys. Rev. {\bf D18} (1978) 2013.}


\end{thebibliography}
\end{document}